# Implementation of quark confinement, and retarded interactions algorithms for Chaos Many-Body Engine

I.V. Grossu[a,*], C. Beşliu[a], Al. Jipa[a], D. Felea[b], E. Stan[b], T. Eşanu[a,c]

[a] *University of Bucharest, Faculty of Physics, Bucharest-Măgurele, P.O. Box MG 11, 077125, Romania*
[b] *Institute of Space Science, Bucharest-Măgurele, P.O. Box MG 23, 077125, Romania*
[c] *„Horia Hulubei" National Institute for R&D in Physics and Nuclear Engineering, IFIN-HH, Bucharest, Măgurele, Romania*

In Grossu et al. (2012) we presented a Chaos Many-Body Engine (CMBE) toy-model for chaos analysis of relativistic nuclear collisions at 4.5 A GeV/c (the SKM 200 collaboration) which was later extended to Cu + Cu collisions at the maximum BNL energy. Inspired by existing quark billiards, the main goal of this work was extending CMBE to partons. Thus, we first implemented a confinement algorithm founded on some intuitive assumptions: 1) the system can be decomposed into a set of two or three-body quark white clusters; 2) the bi-particle force is limited to the domain of each cluster; 3) the physical solution conforms to the minimum potential energy requirement. Color conservation was also treated as part of the reactions logic module. As an example of use, we proposed a toy-model for p + p collisions at $\sqrt{s}$ = 10 GeV and we compared it with HIJING. Another direction of interest was related to retarded interactions. Following this purpose, we implemented an Euler retarded algorithm and we tested it on a simple two-body system with attractive inverse-square-law force. First results suggest that retarded interactions may contribute to the Virial theorem anomalies (dark matter) encountered for gravitational systems (e.g. clusters of galaxies). On the other hand, the time reverse functionality implemented in CMBE v03 could be used together with retardation for analyzing the Loschmidt paradox. Regarding the application design, it is important to mention the code was refactored to SOLID. In this context, we have also written more than one hundred unit and integration tests, which represent an important indicator of application logic validity.

## 1. Introduction

Inspired by existing studies on Fermi nuclear systems [1-4], in a previous Chaos Many-Body Engine (CMBE) version [5], we proposed a toy-model for relativistic nuclear collisions at 4.5 A GeV/c, and we tested it on experimental data from SKM 200 collaboration [6-9]. In [10] we extended the model to Cu + Cu collisions at 200 A GeV/c (the maximum BNL energy). A HIJING-CMBE comparative study was also discussed in this context. Starting from the existing quark billiards [11], the main goal of this paper was to extend CMBE to partons. One of the most important computational difficulties we faced in this attempt is related to color confinement [12]. For simplicity, we chose an algorithm based on some intuitive assumptions [13]: 1) the system can be decomposed into a set of two or three-body quark white clusters; 2) the bi-particle force is limited to the domain of each cluster; 3) the physical solution conforms to the minimum potential energy requirement. Color conservation was also treated as part of the reactions engine module. As an example of use, we implemented a toy-model for p + p collisions at $\sqrt{s}$ = 10 GeV.

Considering the importance of interaction retardation in relativistic many-body systems [14], we have also implemented an Euler retarded algorithm and we applied it on a two-body system with attractive inverse-square-law force. First results suggest that retarded interactions may contribute to the Virial theorem anomalies encountered for gravitational systems (e.g. clusters of galaxies) [15]. On the other hand, the time reverse functionality implemented in CMBE v03 [16] could be used together with retardation for analyzing the Loschmidt paradox [14].

Taking into account the new features complexity, refactoring CMBE to SOLID [17] proved to be an important prerequisite. Unit testing [18] is another direct benefit of the new application design. We thus provided more than one hundred unit and integration tests covering the most important functionalities from all application logic modules.

## 2. Program description

Our initial goal was to create a high flexible, object oriented C#. Net solution for chaos analysis of relativistic many-body systems. With complexity growth, refactoring CMBE to SOLID [17] proved to be a critical requirement. As the code was discussed in detail in some previous papers [5,10,16,19], in this work we will focus on the most important changes.

**The entity and data access layers (*Engine.Data.dll*)**. As a first refactoring step, we tried to separate data from logic by creating specific entities: *Particle, NBody, QCDColor, Reaction* etc. The basic functionalities for accessing comma separated values files are included in the *Data.dll* library, while the responsibility of persisting entities is granted to dedicated classes, each implementing a corresponding interface (e.g. *INBodyData, IQcdNBodyData* etc.) in agreement with SOLID. It is important to note that the file structure used for storing simulations outputs (header and data csv files) is not backward compatible. Thus, the old *Id* column was renamed to *Code* and represents the particle type defined in any convenient schema (e.g. PDG), while the *Id* column is now an integer uniquely identifying each particle inside its parent many-body system (relational model). Information on force acting on each particle was also added to the data csv file. For implementing reaction schemas, we are using the *DsReaction* typed dataset together with the *ReactionData* data access class. For performance reasons, the available decays and bi-particle reactions are loaded in two generic dictionary objects (*ReactionData.Decays*, and *ReactionData.Reactions*). A new property (*Class*) was added for storing information on particle category (quark, gluon etc.). Starting with the current version one can provide also mass ranges instead of simple values (the *MassWidth* property). A new string property (*Tag*) was added to the *Reaction* table. When not null it could be used for selecting reactions according to specific criteria available at runtime (see, for example, the *QcdNBodyRungeKuttaAlgorithm class, GetDecayTag()* method, where the *Tag* property was used to communicate to Reaction Engine the constituents number of the cluster to which the current quark belongs).

**The application logic layer (*Engine.dll*)**. We started refactoring the application logic by separating the Runge-Kutta simulation algorithm into a set of loosely coupled classes. Thus, we gain the possibility of dynamically injecting specific behaviors (the strategies design pattern [17]).

The *RawNBodyEulerAlgorithm* class implements the *IRawNBodyAlgorithm*. It provides an Euler algorithm for computing the equations derived from the second Newton's law for a relativistic many-body system. *NBodyEulerAlgorithm* extends it for adding calculation of both Virial Coefficient and energy conservation test [20].

The *NBodyRungeKuttaAlgorithm* reuses the *NBodyEulerAlgorithm* logic (inheritance) for implementing a second order Runge-Kutta algorithm. It is loosely coupled with an *IRawNBodyAlgorithm* (the *IntermediaryNBodyAlgorithm* property) which default's value is an Euler algorithm instance. The "intermediary" algorithm's responsibility is to calculate the bi-particle force for the midpoint of the integration interval *h*. Thus, the described mechanism is equivalent with the initial, less structured, second order Runge-Kutta implementation. Although obsolete, the old code was kept (the *NBodyDefaultAlgorithm* class) just for historical reasons and as a unit tests reference.

The *NBodyEngine* class is now the many-body simulation "orchestrator". It is loosely coupled with all necessary components: the data access class used for persisting the simulation output (the *INBodyData* interface), the many-body simulation algorithm (*INBodyAlgorithm*), and the reaction engine described in [5], (*IReactionEngine*). One of the main benefits of this approach is the gain in flexibility. On the other hand, it is a unit testing prerequisite, allowing substitution of dependent components with „mocks" [18].

The *Universe* class contains the main loop for simulating a collection of many-body systems. It provides an elegant "time generator", based on the *yield return* C# statement [21]. The time generator has also a reverse version which, together with interaction retardation, could be used for analyzing the Loschmidt paradox [14]. *Universe* implements the *IUniverse* interface, which in turn implements *IParallelObject*, thus facilitating parallelization (see the thread-based framework implemented in *PrallelProgramming.dll*).

*SimulationBase* is the base class to be used for each specific simulation of interest (the template design pattern). It is now loosely coupled with *Universe*.

A similar refactoring was considered for the CMBE high precision framework, too [19].

### 3. Implementation of a quark confinement algorithm

For implementing color confinement, we considered some intuitive hypotheses:
- although gluons are not explicitly implemented, their kinetic effect is included in quark masses (the constituent model [22,23]);
- each system of interest can be decomposed into a collection of two or three-body quark white sub-systems (clusters);
- the bi-particle interaction is restricted to the scope of each white cluster;
- from all possible cluster configurations, the physical solution must conform to the minimum potential energy requirement [13].

We have first designed some entities for employing the color property. Thus, *QcdColorComponent* has two integer properties *Value,* and *AntiValue* (this will facilitate gluons implementation). *QcdColor* encapsulates a three-component array (*Red, Green, Blue*). Together with various specific functionalities it provides also an overloaded version of the plus operator. A *QcdColor* property was added to the *Particle* class. The *QcdBag* entity is used for implementing systems of colored objects, while *QcdBagSystem* is a collection of *QcdBag* objects. It also encapsulates a generic dictionary for establishing if two distinct quarks are members of the same bag.

The *IQcdConfinementAlgorithm* interface is the contract to be implemented by any QCD confinement algorithm. The *QcdQuarkMax3Confinement* class provides a recursive backtracking algorithm for searching the white clusters configuration satisfying the minimum potential energy requirement (the *SearchOptimalBagSystem* method).

The *QcdRawNBodyEulerAlgorithm* is implementing an Euler algorithm for quark many-body systems with bag scope interaction (see the overridden versions of *GetBiparticleForce*, and *OnBeginComputeBiparticleParameters* methods).

It inherits the *RawNBodyEulerAlgorithm* and is loosely coupled with a confinement algorithm. The *QcdNBodyRungeKuttaAlgorithm* class inherits *NBodyRungeKuttaAlgorithm* and is mainly used for replacing the default intermediary Euler algorithm with a *QcdRawNBodyEulerAlgorithm* instance.

The reaction engine is loosely coupled with the *QcdReactionLogic* class. In this version we considered only reactions in which either the reacting particles colors are randomly redistributed between the reaction products or a white color/anti-color pair is generated.

As an example of use, we implemented a quark toy-model for p + p collisions at relativistic energies (*SimulationQuarkCollisionExample* class). For simplicity, we chose harmonic interactions [24]. The quark constituent masses were set to the traditional values of $m_u=m_d≈300MeV$, and $m_s=486MeV$. Each proton is composed by three quarks (*u,u,d*), with colors (*R,G,B*) assigned in an arbitrary order. The quark constituents are initially placed in the vertices of a randomly oriented equilateral triangle, rotating around the system's geometric center. The total kinetic energy ($E_k≈38MeV$) was set to the difference between hadron's and its constituents' masses, while the triangle radius $r_Δ$, and the quark-quark interaction distance $d_σ$ were chosen in agreement with proton's radius ($r_Δ+d_σ≈1.15Fm$). The available energy in the center of mass system is a simulation parameter.

A set of $q_1 + q_2 → q_1 + q_2 + q + \bar{q}$ pair-generation reactions were considered (see the *QuarkReactions.xml* file). One important difficulty we faced in this context is related to the fact that, as result of applying the confinement algorithm, the reactions products could be captured in 2-body systems. In absence of a particle recombination algorithm this would result in overestimating Mesons' masses. As a workaround solution, we employed two quark categories: "hadron quarks" (*Class=„Quark", Mass=300MeV*), and „meson quarks" (*Class=„Quark", Mass=60MeV*). One hadron quark will decay into a meson quark and a photon, after an enough long time (*MeanFreeTime=35Fm/c*), only if it is part of a 2-body cluster (*Reaction.Tag="Meson"*, see the, previously discussed, *QcdNBodyRungeKuttaAlgorithm* class, and the runtime reactions filtering mechanism).

HIJING [25,26] is a Monte Carlo program based on QCD models for multiple jets production. In Fig.1 we compared HIJING and CMBE hadron distributions for 10000 p + p events at $\sqrt{s}$ = 10 GeV, with collision parameter in [0,1] Fm, and temporal resolution $h=10^{-2}$ *Fm/c* (the Energy being conserved, in this case, with $10^{-2}$ precision). The observed differences could be only in part explained by the fact CMBE is missing both a particle recombination mechanism and an exhaustive reactions schema (e.g. CMBE's output does not contain Electrons). However, considering all discussed simplifications, the proposed toy-model could be understood as a first promising step in our attempt.

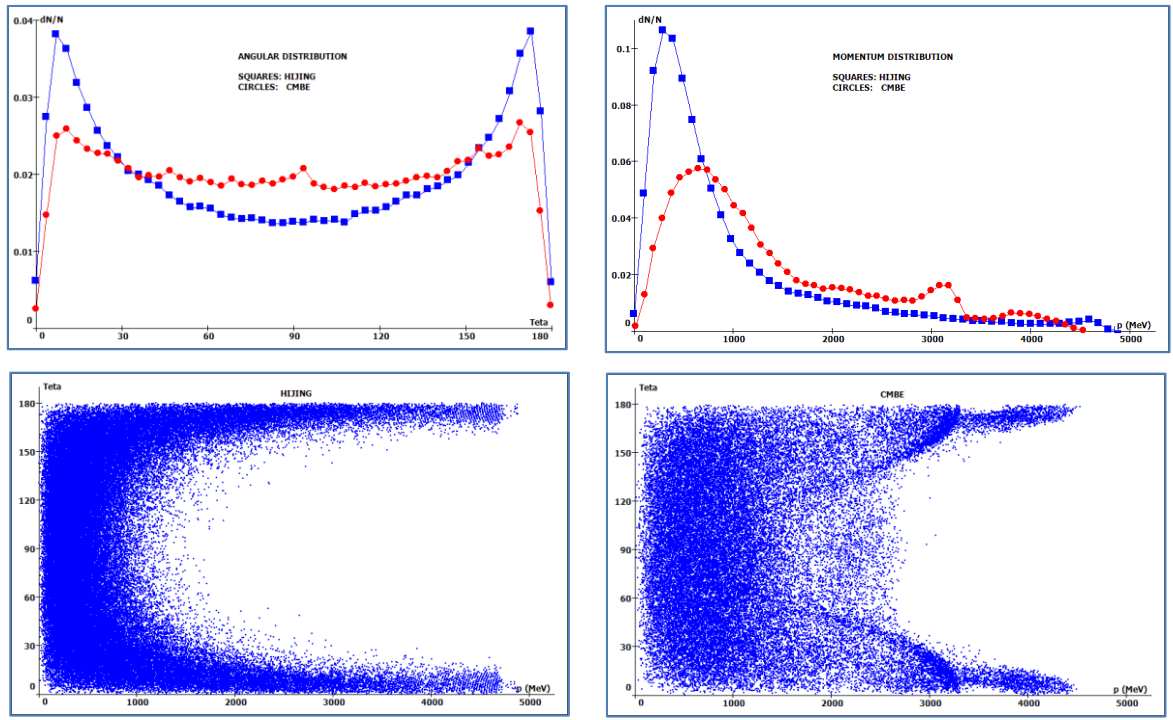

**Fig.2.** Comparison between CMBE and HIJING hadron distributions for p + p at $\sqrt{s}$ = 10 GeV. Up left: Angular distribution. Up right: Momentum distribution. Down left: HIJING (theta, p). Down right: CMBE (theta, p).

## 4. Implementation of an interaction retardation algorithm

Our first attempt of implementing retarded interactions is described in [27]. The application SOLID design is now facilitating this goal. Thus, the *RetardedNBodyEulerAlgorithm* class inherits *NBodyEulerAlgorithm* to add interaction retardation logic to the Euler algorithm. The mechanism is based on storing the system evolution in time into a queue data structure (the *History* property). For avoiding memory consumption issues, the queue capacity is limited to a convenient value provided as a constructor parameter. As the third Newton's law is not valid for systems with retarded interactions [28], the *ComputeParticlePairs* method was overridden for generating particle 2-permutations instead of 2-combinations. The interaction speed $c_i$ is specified as a constructor parameter. Each particle is supposed to interact only with the historical instances for which the delay criterion is satisfied: $\Delta r = c_i * \Delta t$ (Greedy programming technique). The *CouldntFindHistoricalParticle* Boolean property is used for indicating any issue in finding an appropriate solution resulted by either expected retardation effects, or queue insufficient capacity (see the *GetRetardedForce* method). The algorithm complexity thus becomes $O(p \times n^2)$, where $p$ is the queue capacity.

As an example of use we considered a simple relativistic two-body system with attractive-inverse-square law force (*SimulationRetardedInteractionExample*). Using an arbitrary system of units, we chose $m_1 = 10000$, $m_2 = 1$, and $F = 10^{-5}$ $m_1$ $m_2$ / $r^2$. The particles are initially at rest, placed at a mutual distance $r_0=1$. Working with a temporal resolution $h=10^{-2}$ the energy is conserved with $10^{-4}$ precision. In this context, we empirically noticed the Virial coefficient (implemented in CMBE v01) is sub-unitary and reaches the expected value (one) as the interaction speed approaches infinity (Fig.2). This suggests that interaction retardation may contribute to the virial theorem anomalies encountered for gravitational systems (e.g. clusters of galaxies) [15]. It is important to mention that values higher than the speed of light were artificially provided only for analyzing the classical limit. For excluding any unwanted effect related, for example, to the queue limited capacity, we also implemented a set of unit tests (see the *RetardedInteractionTests* class) for checking, at each iteration, the previously described *CouldntFindHistoricalParticle* property, and the Virial Coefficient values.

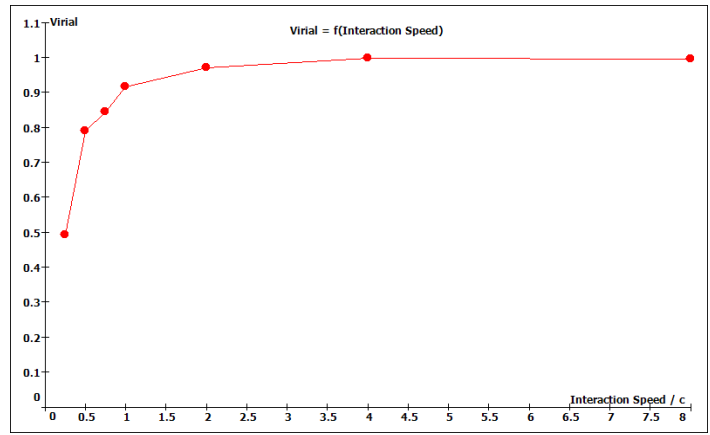

**Fig.2** The Virial Coefficient for a two-body gravitational-like system, represented as a function of interaction speed. It is important to mention that artificial values, higher than the speed of light, were provided only for analyzing the classical limit (infinite interaction speed).

## 5. Conclusion

Inspired by existing quark billiards [11], the main goal of this work was extending CMBE [10] to partons. In this context, we implemented a color confinement algorithm and we tested it on a toy-model for p + p collisions at $\sqrt{s}$ = 10 GeV. Bearing in mind all previously discussed simplifications, the relative agreement CMBE - HIJING presented in (Fig.1) is validating the model as a first encouraging step in our attempt.

Considering the importance of interaction retardation in relativistic many-body systems [14, 28], we also implemented an Euler retarded algorithm and applied it to a simplified two-body system with attractive inverse-square-law force. In this context we empirically noticed (Fig.2) that the Virial Coefficient reaches the expected value (one) only in the classical limit (infinite interaction speed). This suggests that retardation may contribute to the virial theorem discrepancies (dark matter) encountered for gravitational systems (e.g. clusters of galaxies) [15].

The previously described examples of use do not require significant resources to run (CPU 1.0GHz, 128M free RAM, .Net Framework 4.0 running on MS Windows XP or later). However, the resources could become a critical problem with increasing of reactions and particle numbers.

Taking into account the growing complexity, in contrast with our initial goal of creating an object oriented, flexible solution, refactoring CMBE to SOLID [17] proved to be a critical requirement. The new application design facilitates implementation of new features and is also a unit testing prerequisite [18]. In this context, we provided more than one hundred unit and integration tests which represent an important application logic quality indicator.

**Acknowledgements**

*This work was supported by 09-FAIR/16.09.2016, FAIR-RO and University of Bucharest project nr. 48/2018*